\documentclass[conference]{IEEEtran}
\usepackage{algpseudocode}
\usepackage{algorithmicx}
\usepackage[ruled,vlined,linesnumbered]{algorithm2e}
\usepackage{amssymb}
\usepackage{booktabs}
\usepackage{graphicx, cite}
\usepackage{subfigure}
\usepackage{amsmath}
\usepackage{amsthm}
\usepackage{float}
\usepackage{stfloats}
\usepackage{url}
\usepackage{multirow}
\usepackage{xcolor}
\usepackage{CJK}
\usepackage{enumitem}
\IEEEoverridecommandlockouts

\def\BibTeX{{\rm B\kern-.05em{\sc i\kern-.025em b}\kern-.08em T\kern-.1667em\lower.7ex\hbox{E}\kern-.125emX}}

\begin{document}
\title{
Visualization of Mobility Digital Twin: Framework Design, Case Study, and Future Challenges}

\author{
\IEEEauthorblockN{Yueyang Liu\IEEEauthorrefmark{1}, Xiaolong Tu\IEEEauthorrefmark{1}, Dawei Chen\IEEEauthorrefmark{2}, Kyungtae Han\IEEEauthorrefmark{2}, Onur Altintas\IEEEauthorrefmark{2}, Haoxin Wang\IEEEauthorrefmark{1},\\ and Jiang Xie\IEEEauthorrefmark{3}\\
\IEEEauthorblockA{\IEEEauthorrefmark{1}\textit{Department of Computer Science, Georgia State University, GA, USA}}
\IEEEauthorblockA{\IEEEauthorrefmark{2}\textit{InfoTech Labs, Toyota Motor North America R\&D, CA, USA}}
\IEEEauthorblockA{\IEEEauthorrefmark{3}\textit{Department of Electrical and Computer Engineering, The University of North Carolina at Charlotte, NC, USA}}}
\thanks{This work was supported in part by the U.S. National Science Foundation (NSF) under Grant No. 1910667, 1910891, 2025284, and funds from Toyota Motor North America.}
}

\maketitle
\thispagestyle{plain}
\pagestyle{plain}
\pagenumbering{gobble}

\begin{abstract}
A Mobility Digital Twin is an emerging implementation of digital twin technology in the transportation domain, which creates digital replicas for various physical mobility entities, such as vehicles, drivers, and pedestrians. Although a few work have investigated the applications of mobility digital twin recently, the extent to which it can facilitate safer autonomous vehicles remains insufficiently explored. In this paper, we first propose visualization of mobility digital twin, which aims to augment the existing perception systems in connected and autonomous vehicles through twinning high-fidelity and manipulable geometry representations for causal traffic participants, such as surrounding pedestrians and vehicles, in the digital space. An end-to-end system framework, including image data crowdsourcing, preprocessing, offloading, and edge-assisted 3D geometry reconstruction, is designed to enable real-world development of the proposed visualization of mobility digital twin. We implement the proposed system framework and conduct a case study to assess the twinning fidelity and physical-to-digital synchronicity within different image sampling scenarios and wireless network conditions. Based on the case study, future challenges of the proposed visualization of mobility digital twin are discussed toward the end of the paper.
\end{abstract}


\section{Introduction}
Technologies that enable connected and autonomous vehicles have been research and development focus of both the academia and the industry. 
In particular, the shift towards \underline{C}onnectivity, \underline{A}utonomous driving, \underline{S}hared mobility, and \underline{E}lectrification of vehicles (CASE) mobility has been the focus of many automotive original equipment manufacturers (OEMs).
For example, in 2019, Toyota announced a profound transformation from being an automaker to becoming a mobility company, with an emphasis on the CASE \cite{ToyotaCASE}.
General Motors, in early 2020, unveiled its new all-electric architecture and emphasized its commitment to autonomous technology, intending to create a zero crashes, zero emissions, and zero congestion future, largely driven by CAVs \cite{GMCASE}.

The emergence of mobility digital twin technology offers new possibilities towards realizing the aforementioned objectives. 
A mobility digital twin aims to create digital replicas for various mobility entities, such as vehicles, drivers, and pedestrians, in the virtual world based on the data acquired in the real world \cite{9724183,wang2023metamobility}.
A few recent studies have envisioned that the mobility digital twin can enhance safety for CAVs. For instance, a digital behavior twin framework has been proposed to model driver behaviors and share them among connected vehicles. The shared digital behavioral models can be used for predicting future actions of neighboring vehicles and improving driving safety \cite{chen2018digital}. 
In addition, a few research work have investigated the deployment of mobility digital twins. For example, a human-in-the-loop simulator that consists of a cloud server, the Unity game engine, and the Logitech G29 Driving Force is developed to emulate the vehicle data sampling in the physical space \cite{wang2021digital}. The sampled driving data can be uploaded to the cloud server for driving behavior learning and modeling. 


Although there have been notable advancements in the development of mobility digital twin technology, its potential for enhancing the safety of CAVs is still largely untapped. 
Specifically, most existing work on mobility digital twins focuses only on sampling and twinning limited information about vehicles, such as their positions, speeds, and accelerations. 
This limits the ability of mobility digital twin comprehensively representing the attributes of mobility entities and provide accurate decisions under complex environmental uncertainties.
Adding geometry representations of mobility entities, such as 3D shape, appearance, and articulated geometry may help to create a comprehensive mobility digital twin
which can significantly improve the perception accuracy. 

In this paper, we propose \textit{visualization of mobility digital twin}, an approach for creating high-fidelity and manipulable geometry representations for physical mobility entities in the digital space. The twinned high-fidelity geometry representations can \textit{augment the existing CAV perception and provide more accurate scene understanding,} particularly in complex environments.
For example, an ego CAV is driving on a highway and encountering a vehicle in front of it with a flat tire. Human drivers might slow down or change lanes to avoid getting too close to the vehicle. However, for a CAV, detecting and understanding this situation is much more challenging. This is where the high-fidelity geometry representations come into play. The twinned high-fidelity geometry representation of the vehicle with the flat tire can provide information about the position, size, and orientation of the flat tire, as well as any other relevant details about the vehicle's state. Using this information, a CAV can make informed decisions about how to respond to the situation.
Besides the perception augmentation, the twinned manipulable geometry representations can \textit{facilitate realistic training and large-scale validation of CAVs in the digital space,} particularly for corner cases. For instance, by manipulating twinned mobility entities (e.g., vehicles and pedestrians), a wide range of new behaviors and emergent situations can be generated, such as adding a child chasing a ball to the driving path \cite{Nvidia-sim}. These benefits demonstrate that the visualization of mobility digital twin can enhance the reliability and explainability of CAV decisions.

However, implementing the visualization of mobility digital twin in the real world is significantly challenging. First, creating high-fidelity and manipulable 3D geometry representations for physical mobility entities is labor-intensive and time-consuming. Conventional methods for creating geometry representation of vehicles require skilled engineers and artists, such as computer-aided design (CAD) models \cite{wang2022cadsim}.
Second, although several machine learning based methods have been developed for creating 3D geometry models \cite{wang2022cadsim,mildenhall2021nerf,song2021vis2mesh}, all these methods require a large amount of data collected in the physical space, such as camera images or LiDAR point clouds.
Third, timely updates on the geometry representations are crucial for reflecting the physical changes in the digital space, but it also presents significant challenges.

To tackle the above challenges, we propose a novel system framework with edge computing to support cost-efficient and scalable visualization of mobility digital twins. The proposed system framework consists of sensory data collection with crowdsourcing, data preprocessing and offloading, and high-fidelity 3D geometry model reconstruction. The novel contributions of this work are summarized as follows:
\begin{itemize}
    \item We design the first edge computing system for the visualization of mobility digital twins based on crowdsourcing and Neural Radiance Fields (NeRF). The designed system can automatically create and update high-fidelity 3D geometry models for mobility entities through camera images crowdsourced by neighboring vehicles.
    \item We prototype the proposed system on a physical testbed, and conduct a case study with extensive performance evaluations in four image sampling scenarios.
    \item We identify potential research challenges that need to be addressed for the development of visualization of mobility digital twins based on our observations from the prototyped system.
\end{itemize}

\section{Background on Visualization of Mobility Digital Twins}
\label{secBackground}
In this section, we primarily examine the rationale behind the need for the visualization of mobility digital twins in the context of CAVs.
Specifically, we discuss the concept of CAVs, mobility digital twin, and the state-of-the-art 3D geometry model reconstruction technologies.

\subsection{Connected and Autonomous Vehicles}
The concept of CAV emerges in the 1980s with the technological advances in sensing, communication, and computing \cite{guanetti2018control}.
CAVs are vehicles that can drive themselves using a combination of advanced control and computer vision technologies and various sensors, such as camera, radar, LiDAR, GPS, and odometry. They are also typically capable of machine learning, meaning they can adapt to changing conditions and improve their performance based on experience.
The ``connected" aspect of CAVs refers to their ability to communicate with other vehicles and infrastructure through Vehicle-to-Vehicle (V2V) and Vehicle-to-Infrastructure (V2I) communications. This allows for more efficient use of roadways, as vehicles can communicate to avoid collisions, navigate around traffic congestion, and better react to changing road conditions.
Existing research envision that the introduction of CAV could make driving safer and more sustainable \cite{ye2019evaluating, wang2020architectural}. 

\subsection{Mobility Digital Twin for CAVs}
The mobility digital twin is a system that implements the concept of digital twin \cite{glaessgen2012digital, jones2020characterising} in the transportation domain. It consists of a digital space, a physical space, and a communication plane for data exchange between those two spaces \cite{9724183,liu2022poster}. Although many studies consider a general digital twin system as a high-fidelity modeling and simulation platform, the mobility digital twin goes beyond simulation. For example, a mobility digital twin can include sampling and actuation in the physical space, and storage, modeling, learning, simulation, and prediction in the digital space.
By leveraging the mobility digital twin, OEMs not only can thoroughly assess their CAV stack in diverse scenarios through an interactive and immersive process, but also improve the overall reliability, safety, and sustainability of CAVs by facilitating more informed decision-making and planning \cite{kuvsic2023digital,fan2021digital}.

\subsection{State-of-the-Art 3D Geometry Model Reconstruction}
The core component of the visualization of mobility digital twin is the 3D geometry model reconstruction, the process of creating a high-fidelity digital 3D representation from a sequence of sensory data collected in the physical space. There are several techniques designed for 3D representations in computer vision (none of them are specifically developed for the visualization of mobility digital twins).

\subsubsection{Neural Radiance Fields (NeRF)}
NeRF is one of the state-of-the-art machine Learning techniques for 3D geometry model reconstruction \cite{mildenhall2021nerf, deng2022depth,mueller2022instant,li2022rt, zhang2021ners,wang2021nerf,xie2023snerf}. It uses multi-layer perceptron (MLP) networks to model the appearance of 3D objects from 2D images. In essence, NeRF learns a continuous 3D representation of an object or scene by processing a large number of 2D images captured from various viewpoints. The deep neural network employed in NeRF maps a 3D location and viewing direction to a radiance value, which is the color of the corresponding point. By taking the 2D images as input, the network generates radiance values for each 3D location and viewing direction, resulting in a highly realistic 3D representation of the object or scene.

\subsubsection{LiDAR Point Cloud}
The point cloud is a 3D data representation that comprises a collection of points in three-dimensional space. Each point in a point cloud represents a particular location in space and may contain additional information such as color, intensity, or other characteristics. Typically, point clouds are generated using 3D scanning technologies such as LiDAR.
As a widely used technology for visualization in 3D models of object creation and environment simulation, there has been significant research in developing algorithms for processing and analyzing point clouds \cite{guo2020deep,bisheng2017progress}. However, advanced LiDAR systems can be expensive, and this would be a barrier to the widespread adoption of LiDAR point cloud-based 3D representations.

\subsubsection{3D Mesh}
The 3D mesh is a classic 3D representation that is widely used in computer graphics, simulation, and visualization \cite{song2021vis2mesh}. It is a surface-based geometry that comprises a collection of vertices, edges, and faces. Vertices are the points in space that define the shape of the object, while edges connect the vertices to form a network of interconnected lines. However, 3D meshes have some limitations, such as difficulty in representing complex shapes and surfaces, and capturing fine details.

\textit{Because of the need for cost-efficiency and high-fidelity in implementing the visualization of mobility digital twin, our proposed work builds upon NeRF.} Since compared to point cloud and 3D mesh-based methods, NeRF can generate a 3D representation for an object or scene with comparable (or even better) fidelity using only limited 2D images.

\begin{figure*}[t]
\centering
\includegraphics[width=1\textwidth]{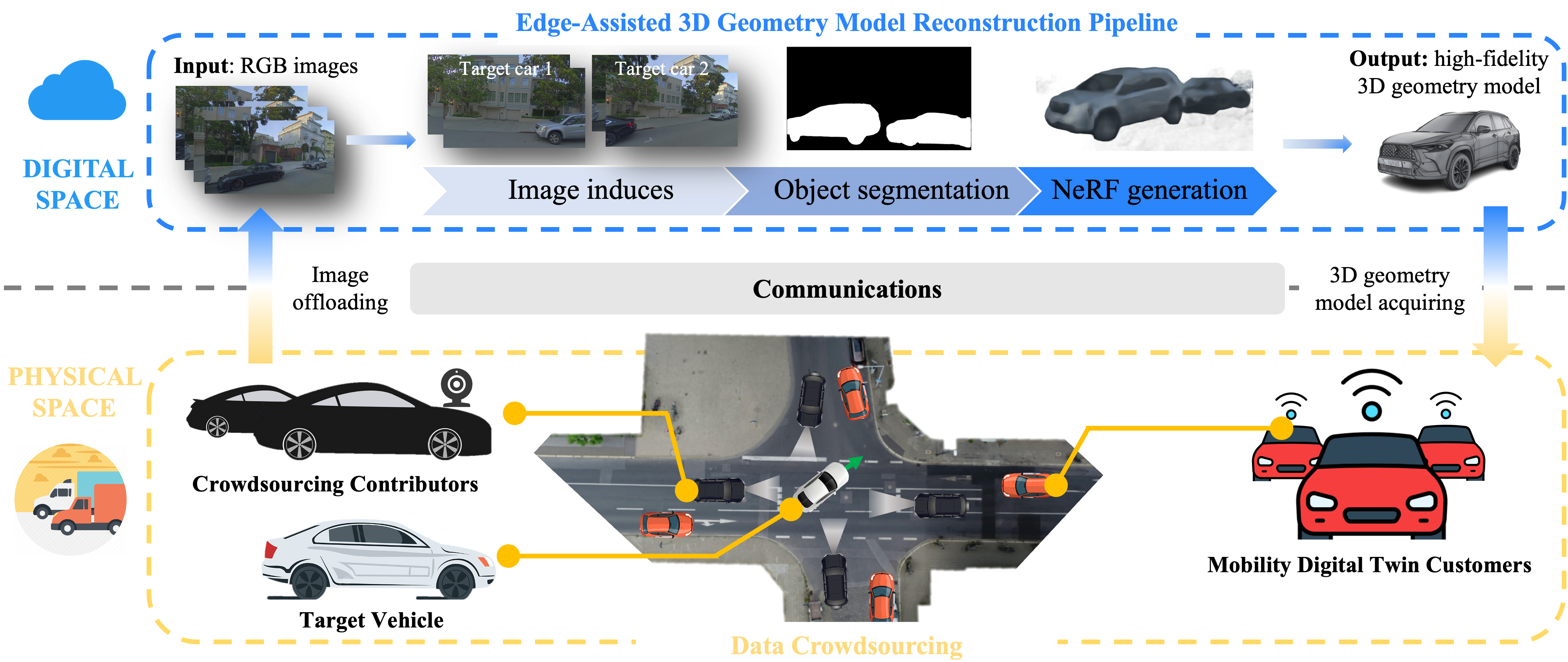}
\centering
\caption{Overview of the proposed system framework for visualization of mobility digital twin with edge computing, which consists of a physical space, a digital space and a communication plane between two spaces.}
\label{fig:structure}
\centering
\end{figure*}

\section{Proposed Framework for Visualization of Mobility Digital Twin}

In this section, we present the overview of our proposed framework for visualization of mobility digital twin. Then, we describe key attributes of the twinned 3D geometry models.

\subsection{System Overview}
Our goal of the visualization of mobility digital twin is to automatically create and update high-fidelity 3D geometry models for physical mobility entities, such as vehicles and pedestrians, in a cost-efficient and scalable manner.
Towards this goal, we propose and design an edge-assisted system for the visualization of mobility digital twins by leveraging crowdsourcing and NeRF. The architecture of the system proposed in this work, as illustrated in Fig. \ref{fig:structure}, consists of three components: the \textit{physical space} where mobility entities such as vehicles and pedestrians reside, the \textit{digital space} where the twinned digital replicas of those physical mobility entities and the edge-assisted 3D geometry model reconstruction pipeline are located at, and the \textit{communication plane} that support data exchange between these two spaces.

\subsubsection{Physical Space}
In the physical space, CAVs can be categorized into three classes: \textit{crowdsourcing contributor, target vehicles,} and \textit{mobility digital twin customers.} Note that a single CAV may belong to multiple categories simultaneously. To address the cost-efficiency and to ensure sufficient image data for 3D geometry model reconstruction, in our proposed framework, the data collection will be performed in a crowdsourcing manner. Thus, the crowdsourcing contributors are responsible for sampling 2D images of the target vehicle (e.g., the white vehicle located at the middle of the intersection in Fig. \ref{fig:structure}) from various camera views. The crowdsourced images will be pre-processed and offloaded to the digital space through the communication plane for 3D geometry model reconstruction. Additionally, the twinned high-fidelity 3D geometry models can be acquired by the mobility digital twin customers to support a variety of vehicular applications, including perception augmentation, which can enhance safety. 

\subsubsection{Digital Space}
The digital space is implemented in edge servers since edge computing offers several advantages for real-time processing and reducing network latency. The edge-assisted 3D geometry model reconstruction pipeline is the core of the digital space, which automatically creates and updates high-fidelity 3D geometry models for target vehicles leveraging a sequence of camera 2D images crowdsourced in physical space. The pipeline includes three key intermediate phases: image induces, object segmentation, and NeRF generation. 

\subsubsection{Communication Plane} In real-world deployment, the communication plane between the physical and digital spaces can be facilitated by various technologies such as 5G, Wi-Fi, or C-V2X.

\subsection{Attributes of 3D Geometry Model}
\label{ssc:attributes}
In this work, fidelity and physical-to-digital synchronicity are the two key attributes of 3D geometry models. These two attributes can be used for the assessment of the overall performance of the visualization of mobility digital twin.

\textit{Fidelity:} 
We define fidelity as the criterion for assessing the precision of the twinned 3D geometry model, without considering the latency-incurred twinning distortion. Essentially, the defined fidelity can purely reflect the impact of the sufficiency and quality of crowdsourced image data, as well as the accuracy of the NeRF generation algorithm, on the precision of the 3D geometry model reconstruction. For example, to achieve a high-fidelity 3D geometry model for a physical mobility entity, a sufficient amount of high-quality image data must be crowdsourced from distinct views and processed by a sophisticated NeRF generation algorithm.

\textit{Physical-to-digital synchronicity:} 
We define physical-to-digital synchronicity as the criterion for measuring the discrepancy between a physical mobility entity and its twinned 3D geometry model, incurred by the end-to-end latency. 
The end-to-end latency, in the proposed framework, includes the time required for crowdsourcing sufficient image data, preprocessing, offloading the preprocessed data from vehicles to the edge server, and processing the edge-assisted 3D geometry model reconstruction pipeline.
Physical-to-digital synchronicity is crucial for timely reflecting and synchronizing the changes in physical mobility entities within the digital space.
Furthermore, it's important to note that the physical-to-digital synchronicity may not always be naively commensurate with the end-to-end latency, as it also depends on how fast and frequent a mobility entity may change under current environmental conditions.


\section{Case Study and Performance Evaluation}
\label{secPrototype And Performance evaluation}
In this section, we implement the proposed framework for visualization of mobility digital twins and conduct a case study of the 3D geometry model reconstruction with image sampling, preprocessing, offloading, and processing. Then, a detailed performance evaluation is conducted to assess fidelity and physical-to-digital synchronicity.

\begin{figure*}[t]
\centering
\includegraphics[width=1\textwidth]{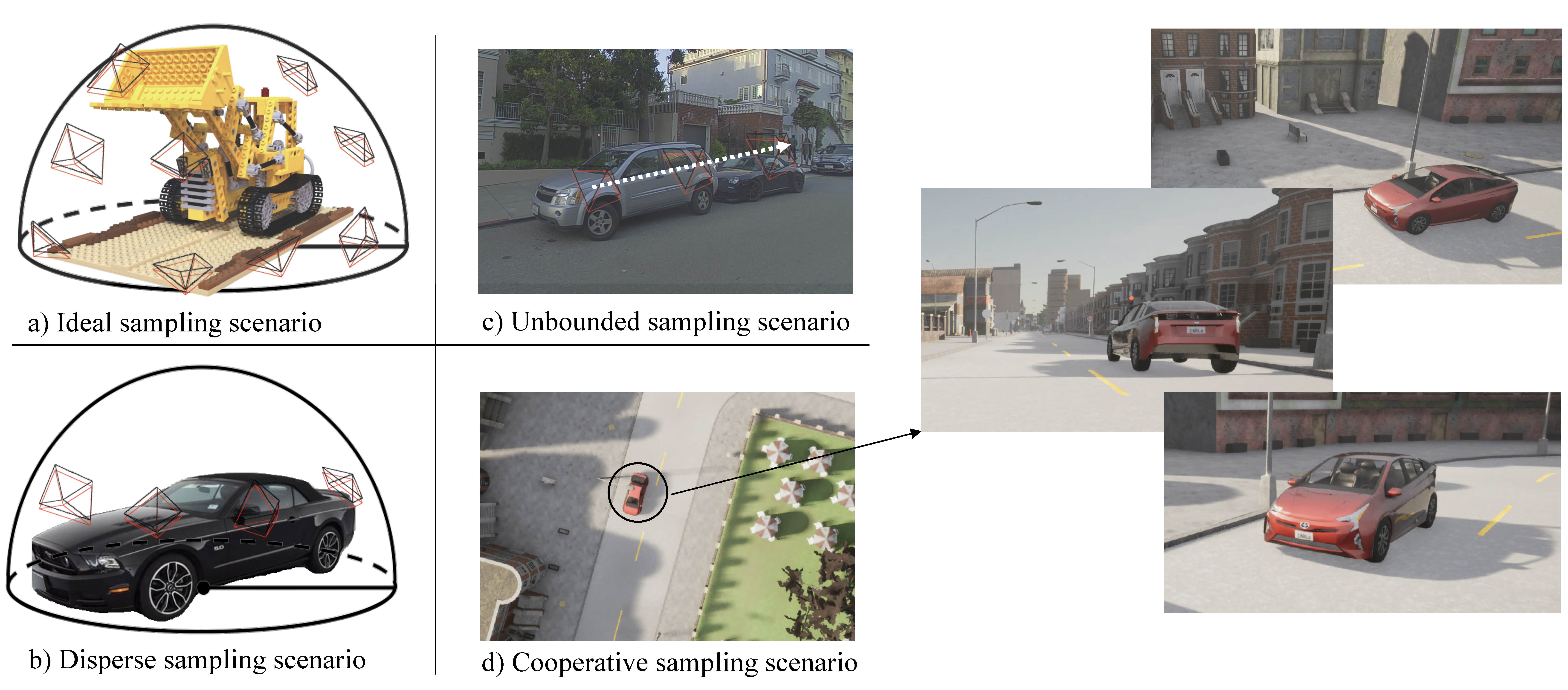}
\centering
\caption{Our proposed four different scenarios for validating the implemented 3D geometry model reconstruction pipeline.}
\label{fig:cav_scenes}
\centering
\end{figure*}

\subsection{System Prototype}
This section presents the prototype of the proposed system framework for the visualization of mobility digital twin, including an edge-assisted 3D geometry model reconstruction pipeline with NeRF, a communication plane with simulated 5G environments, and image datasets.

\subsubsection{Edge-Assisted 3D Geometry Model Reconstruction}
We implement our proposed edge-assisted 3D geometry model reconstruction pipeline, as shown in Fig. \ref{fig:structure}, on an edge server with a single NVIDIA RTX 3090 GPU. 
\textcolor{black}{The implemented 3D geometry model reconstruction pipeline works as follows.
In the initial step of the process, known as image induction, the images captured by each CAV are classified and labeled accordingly to identify their respective vehicles. 
Afterward, the process involves applying object segmentation, which adds a mask to the background object. This step improves the recognition and isolation of the primary vehicle, where Detectron 2, a framework developed by Facebook AI Research, is utilized for this purpose.
}
Furthermore, we implement our NeRF generation mechanism upon a state-of-the-art algorithm, named NVIDIA instant NeRF \cite{mueller2022instant}.
Compared to a conventional 3D mesh-based method that relies on stereo camera images to generate intermediate disparity maps and point clouds, our developed NeRF-based method significantly simplifies the process and does not require stereo camera images as input.

In addition, one of the limitations of NeRF is that it assumes the world is static in terms of geometry, materials, and photometry. This assumption can be severely violated in many real-world scenarios \cite{martinbrualla2021nerf}, especially in the transportation domain. For example, vehicles and pedestrians may move; weather may change; and the sun may move through the sky. Meanwhile, the camera viewpoints can be restricted by the vehicle's trajectory or road topology in the real world, resulting in a limited number of image samples on the target vehicle. Therefore, we create four scenarios, depending on the dynamics of the scene and the number of available sample views, to validate the implemented 3D geometry model reconstruction pipeline, as shown in Fig. \ref{fig:cav_scenes}. The details of those four scenarios are elaborated as follows:
\begin{itemize} 
\item Ideal sampling scenario: In this scenario, the target physical entity is static and small enough, such as a toy car, that a sufficient number of images (e.g., 100 images sampled at distinct viewpoints) can be sampled on a surrounding hemisphere whose center coincides with the entity's center. This is an ideal and common scenario for NeRF 3D geometry reconstruction, where NeRF has been demonstrated to work well in this scenario \cite{mildenhall2021nerf}. However, this scenario is rare in realistic transportation domain.

\item Disperse sampling scenario: In this scenario, the target physical entity is static but large. Images are sparsely sampled on a surrounding hemisphere (e.g., 25 images sampled at distinct viewpoints), since partial viewpoints may not be reached, such as on top of the vehicle. This scenario is more realistic than the ideal sampling scenario because it takes into account the fact that it may not be possible to capture a complete set of images of the physical entity from all viewpoints.

\item Unbounded sampling scenario (street view): In this scenario, the target physical entity is static but large.
However, unlike the above two scenarios, unbounded sampling scenario does not sample images on a hemisphere. The images are sparsely sampled along an unbounded line, which is typically the trajectory of the vehicle that is capturing the images. The trajectory may be restricted by the road topology in the real world. The displacement between the sampled images may be significant, depending on the camera sampling rate and the vehicle's moving speed. In practice, this scenario may occur when the target vehicle is parked at the side of the road, and crowdsourcing contributors are sampling images while moving.

\item Cooperative sampling scenario: In this scenario, the target physical entity is moving and large.
This scenario is an extension of the unbounded sampling scenario. Images are sparsely sampled by multiple neighboring vehicles in the area cooperatively, where the sampled image set may include a variety of perspectives and divergence radii. 

\end{itemize}

\subsubsection{Communication Plane}
\begin{figure*}[t]
\centering
\includegraphics[width=1\linewidth]{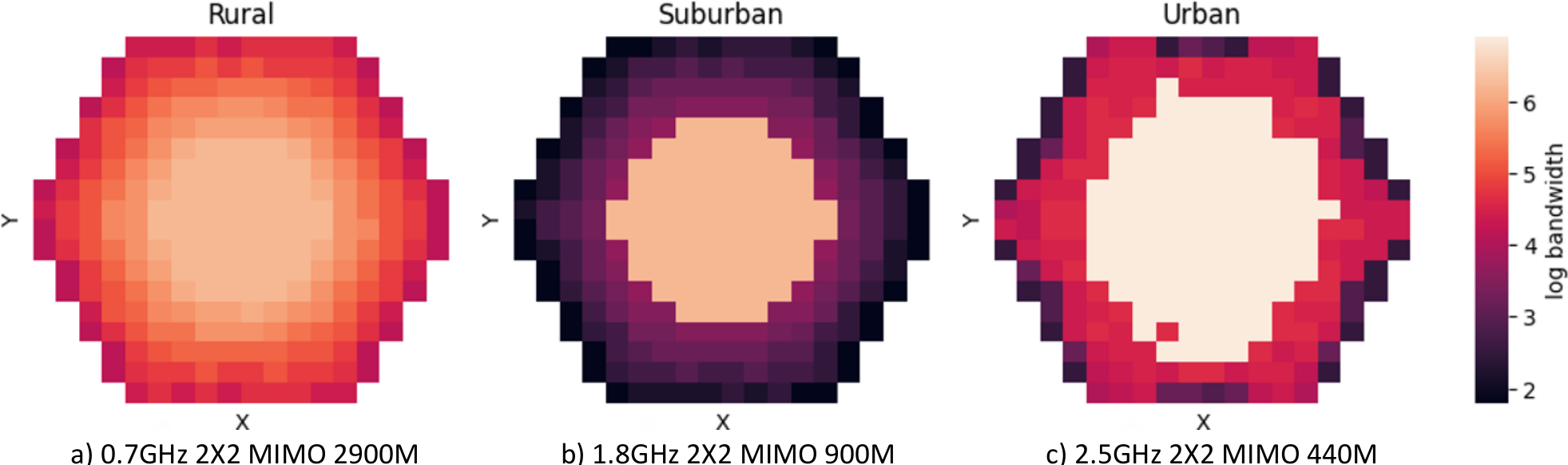}
\centering
\caption{Heatmaps of the simulated 5G network throughput in rural, suburban, and urban areas.}
\label{fig:5gout}
\centering
\end{figure*}

We implement the communication plane on an open-source 5G simulator \cite{oughton2019open}. We simulate three types of 5G network environments:
\begin{itemize}
    \item Rural area: $0.7$ GHz with a 2x2 MIMO, where the distance between two base stations is $2900$ meters.
    \item Suburban area: $1.8$ GHz with a 2x2 MIMO, where the distance between two base stations is $900$ meters. 
    \item Urban area: $2.5$ GHz with a 2x2 MIMO, where the distance between two base stations is $440$ meters. 
\end{itemize}
The simulated 5G network throughput at different locations is illustrated in Fig. \ref{fig:5gout}, which is the input for simulating the vehicle's image offloading and calculating the data transmission latency between the physical and digital spaces (e.g., vehicles and the edge server). We assume that all vehicles associated with the same base station equally share the total bandwidth for the sake of simplicity. In this case, the image offloading process can be simulated by calculating the wireless network throughput based on each vehicle's current location and reducing its remaining data size for offloading accordingly. 
Note that before offloading, the images will undergo pre-processing through compression to conserve network bandwidth. This compression step is implemented to reduce the size of the images, allowing for efficient transfer over the network while preserving essential visual information.

\subsubsection{Image Datasets}
We simplify the implementation of the data crowdsourcing by leveraging three existing image datasets for vehicles. 
\begin{itemize}
\item NeRF synthetic dataset \cite{mildenhall2021nerf}: The NeRF synthetic dataset is used to simulate the ideal sampling scenario. It has $100$ highly overlapped images, where all viewpoints are uniformly sampled on the upper hemisphere.
\item Multi-View Marketplace Cars (MVMC) dataset \cite{zhang2021ners}: The MVMC dataset is used to simulate the disperse sampling scenario. The images are gathered from an online marketplace that hosts thousands of car listings. Each user-submitted listing contains seller images of the same vehicle instance sampled from different perspectives, with an average of $9.1$ images per listing. 
\item PandaSet dataset \cite{xiao2021pandaset}: PandaSet is used to simulate the unbounded sampling scenario, which is a dataset captured by a self-driving vehicle platform equipped with six cameras (left, front left, front, front right, right and back cameras) and two LiDARs. 
\textcolor{black}{
Since the dataset does not provide a camera matrix compatible with NeRF, denoted as $M$, we conduct a matrix transformation through $M=T\times R\times S$.
$R$ is the rotation matrix associated with the given quaternion and is calculated by
$$
\begin{small}
R=\left(\begin{array}{ccc}
1-2 y^2-2 z^2 & 2 x y-2 z w & 2 x z+2 y w \\
2 x y+2 z w & 1-2 x^2-2 z^2 & 2 y z-2 x w \\
2 x z-2 y w & 2 y z+2 x w & 1-2 x^2-2 y^2
\end{array}\right).
\end{small}
$$
$T$ is the translation matrix, calculated by
$$
\begin{small}
T=\left(\begin{array}{cccc}
1 & 0 & 0 & t_x \\
0 & 1 & 0 & t_y \\
0 & 0 & 1 & t_z \\
0 & 0 & 0 & 1
\end{array}\right);
\end{small}
$$
and $S$ is the scaling matrix, calculated by
$$
\begin{small}
S=\left(\begin{array}{cccc}
S_x & 0 & 0 & 0 \\
0 & S_y & 0 & 0 \\
0 & 0 & S_z & 0 \\
0 & 0 & 0 & 1
\end{array}\right).
\end{small}
$$
Additionally, we select the same vehicles (seq id: $30/120/139$) in CADSim \cite{wang2022cadsim} for 3D geometry reconstruction with the left or front-left camera images.
}

\end{itemize}

\begin{figure*}[t]
\centering
\subfigure[Ideal sampling scenario]
{\includegraphics[width=0.25\linewidth]{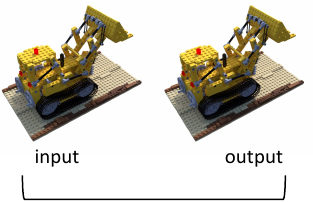}
}
\subfigure[Disperse sampling scenario]
{\includegraphics[width=0.34\linewidth]{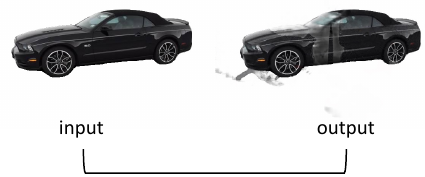}
}
\subfigure[Unbounded sampling scenario]
{\includegraphics[width=0.34\linewidth]{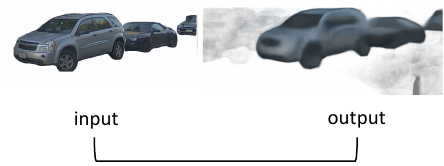}
}
\caption{Qualitative comparison of 3D geometry model reconstruction in different sampling scenarios with NeRF synthetic, MVMC, and PandaSet datasets. 
}
\label{fig:meshoutcome}
\centering
\end{figure*}

\subsection{Performance Evaluation}

\subsubsection{Fidelity} 
In this work, \textcolor{black}{Instant-NGP and NeUS} \cite{mueller2022instant} work as the backbone of the NeRF generation. Fig. \ref{fig:meshoutcome} provides a qualitative comparison among three implemented scenarios, which demonstrates that the ideal sampling scenario achieves the highest visualization quality. 
This is due to the relatively limited number of available perspectives and a high variance radius in disperse and unbounded sampling scenarios, which can lead to a poorer reconstruction characterized by blurs, floaters, and other artifacts in unbounded synthesis.
We also conduct a quantitative comparison among those three scenarios through peak signal-to-noise ratio (PSNR), 
$\text{PSNR}=10\log_{10}(\frac {(L-1)^ {2}}{\text{MSE}})$.
$L$ is the number of maximum possible intensity levels in an image, and MSE refers to mean squared error.
A higher PSNR value indicates a lower level of distortion and a higher fidelity. It is important to note that PSNR is particularly useful for evaluating the quality of lossy compression algorithms, as it focuses on the pixel-wise differences between the original and reconstructed signals.
\begin{table}[t]
\centering
\caption{Quantitative comparison of different reconstruction scenarios in terms of fidelity.}
\begin{tabular}{|c|c|c|c|}
\hline
Dataset & w/o mask & Image number & PSNR \\ \hline
NeRF Synthetic & \checkmark & 100 & 36.79 \\ \hline
MVMC &  & 16 & 34.37 \\ \hline
MVMC & \checkmark & 16 & 37.89 \\ \hline
Pandaset &  & 45 & 14.80 \\ \hline
Pandaset & \checkmark & 45 & 17.25 \\ \hline
\end{tabular}
\label{tb:psnr}
\end{table}

\textcolor{black}{
\textit{Observations and implications:}
As shown in Table \ref{tb:psnr}, 
in comparison to the ideal and dispersed sampling scenarios, the fidelity achieved in the unbounded sampling scenario decreases by $53.1\%$ and $54.4\%$, respectively.
This observation implies that in non-ideal environments, NeRF-based methods may fail to achieve a high fidelity geometry reconstruction for mobility digital twins.
In addition, the datasets processed with the inclusion of a mask achieve higher PSNR values, which advocates selectively highlighting and preserving the visual information of the physical object for enhancing fidelity. The utilization of an opacity mask enables the prioritization and emphasis of the physical object, thereby aiding in mitigating potential distortions or noise introduced during the reconstruction process.
}
\subsubsection{Physical-to-Digital Synchronicity}
The latency of data offloading has a dominant impact on the physical-to-digital synchronicity. Hence, we evaluate the physical-to-digital synchronicity in terms of the offloading latency.
We simulate 5G environments of rural, suburban, and urban areas, where their network throughput is illustrated in Fig. \ref{fig:5gout}.
\textcolor{black}{
Based on the achieved throughput, we evaluate the offloading latency by considering the mobility of vehicles (i.e., generated by a random path algorithm) and the compression of image data.
Observations and analysis of the results are provided in the following section.}

\begin{figure}[t]
    \centering
    \includegraphics[width=\linewidth,trim={0cm 0cm 0cm 0cm},clip]{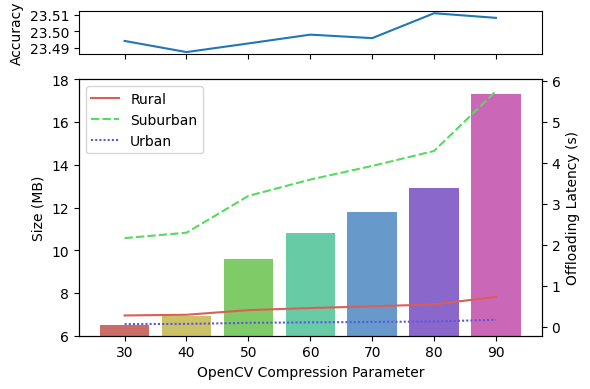}
    \caption{Comparison of image compression ratio, data size, offloading latency, and fidelity.}
    \label{Fig:acc_psnr}
\end{figure}

\subsubsection{Fidelity vs. Physical-to-Digital Synchronicity}
Fig. \ref{Fig:acc_psnr} illustrates the image data size, offloading latency, and fidelity (i.e., accuracy) with different data compression ratios.
The bar chart depicts the relationship between the size of image data for offloading and various compression parameters. The tested compression parameters range from $90$ to $30$, where a lower value indicates a higher compression ratio and a smaller data size. In our experiments, the data size is compressed from over $18$MB to approximately $6$MB.
The line chart compares the offloading latency in urban, suburban, and rural areas with various data sizes.
Urban and rural areas exhibit much lower latency compared to the suburban area. One possible reason for this could be the density and availability of network infrastructure. In the simulated 5G environments, urban areas often have a higher density of network resources and high-frequency wireless networks, which can result in faster data transmission.
Furthermore, in rural areas, there tend to be fewer network users, which leads to less network congestion and, consequently, lower latency.

\textit{Observations and implications:} An interesting observation from our experiment is that the achieved PSNR values (representing fidelity) of the reconstructed 3D geometry model do not vary significantly with respect to the the degree of image compression.
As shown in Fig. \ref{Fig:acc_psnr}, despite the offloaded image being compressed with the lowest compression parameter (i.e., $30$), its achieved PSNR remains relatively close to that of the original image, degrading by only less than $0.1\%$.
\textit{This observation inspires us to adapt the image compression for the physical-to-digital synchronicity improvement by trading as little decrease of the fidelity as possible during the 3D geometry model reconstruction with edge computing.}

\section{Future Challenges}
\label{sc:challenge}

\subsection{Smart Crowdsourcing Contributor Selection}
The evaluation results presented in Table \ref{tb:psnr} and Fig. \ref{Fig:acc_psnr} indicate that the fidelity of the 3D geometry model reconstruction is greatly influenced by the image capture process, particularly the number of unique viewpoints, camera trajectory, and displacement between each image sample. This effect can become particularly important in the cooperative sampling scenarios, since different road topology and traffic conditions may result in varying crowdsourcing performance, such as differences in viewpoint availability and delays.
For instance, it is intuitive that the performance of image crowdsourcing would vary on highways, intersections, and roundabouts. In comparison to intersections and roundabouts, highways have a lower probability of capturing views of the side of the target vehicle. This can result in longer crowdsourcing delays and poorer fidelity in reconstructing the 3D geometry model.
On the other hand, in the case of roundabouts, partial or heavy occlusions are common occurrences. These occlusions can result in only a few qualified viewpoints available for image capturing.
Therefore, smart crowdsourcing contributor selection (i.e., vehicles) is crucial for ensuring both fidelity and physical-to-digital synchronicity but also challenging, especially in cooperative sampling scenarios.

\subsection{Antiquated Segment Detection}
In automated driving, the real-time state of causal traffic participants, such as surrounding pedestrians and vehicles, is the foundation to ensure the autonomy system can act appropriately in various roadway situations.
Hence, it is imperative to accurately detect and update the antiquated segments of the constructed 3D geometry models in the digital space in a timely manner.
The key challenge lies in the obsolete detection and trigger strategy design, because there is no ground-truth for the detection and frequent inferior updates will further exacerbate the network burden. 

\subsection{Fidelity Adaptation in Visualization for Mobility Digital Twin}
In real-world deployment, our proposed framework for visualization of mobility digital twin is expected to scale to the community level, incorporating hundreds of vehicles, while adapting to volatile environmental variabilities.
Hence, how to make the end-to-end process, including image data crowdsourcing, preprocessing, offloading, and 3D geometry model reconstruction, sufficiently scalable with fulfilled fidelity and physical-to-digital synchronicity is the crux of a successful real-world deployment. One of the potential solutions to enhance the system scalability is to intelligently adapt the fidelity of model reconstruction. The adaptation can be based on the current network and computing resources, as well as the fidelity demands of different CAV applications and driving conditions, such as lighting and weather. However, fundamental questions such as how to analytically model the relationship between fidelity and resource allocation; and what are the fidelity demands within different CAV applications and driving conditions still need to be addressed.

\section{Conclusion}
\label{sc:conclusion}
In this paper, we proposed a system framework for the visualization of mobility digital twin with edge computing. To the best of our knowledge, this is the first work that systematically investigates the end-to-end process of constructing 3D visualizations for physical mobility entities in the digital space, which consists of image crowdsourcing, pre-processing, offloading, and 3D geometry model reconstruction with NeRF. We implemented the proposed framework and conducted a case study to evaluate the model fidelity and physical-to-digital synchronicity within different sampling scenarios. The case study showcased the effectiveness of the proposed system framework and uncovered several potential research opportunities to further enhance the system scalability, reliability, and efficiency.

\bibliographystyle{IEEEtran}
\bibliography{ref}

\begin{thebibliography}{10}
\providecommand{\url}[1]{#1}
\csname url@samestyle\endcsname
\providecommand{\newblock}{\relax}
\providecommand{\bibinfo}[2]{#2}
\providecommand{\BIBentrySTDinterwordspacing}{\spaceskip=0pt\relax}
\providecommand{\BIBentryALTinterwordstretchfactor}{4}
\providecommand{\BIBentryALTinterwordspacing}{\spaceskip=\fontdimen2\font plus
\BIBentryALTinterwordstretchfactor\fontdimen3\font minus
  \fontdimen4\font\relax}
\providecommand{\BIBforeignlanguage}[2]{{%
\expandafter\ifx\csname l@#1\endcsname\relax
\typeout{** WARNING: IEEEtran.bst: No hyphenation pattern has been}%
\typeout{** loaded for the language `#1'. Using the pattern for}%
\typeout{** the default language instead.}%
\else
\language=\csname l@#1\endcsname
\fi
#2}}
\providecommand{\BIBdecl}{\relax}
\BIBdecl

\bibitem{ToyotaCASE}
\BIBentryALTinterwordspacing
Toyota, ``Reforming our company to become a ``mobility company",'' Accessed:
  2023-4-26. [Online]. Available:
  \url{https://global.toyota/en/company/messages-from-executives/details/}
\BIBentrySTDinterwordspacing

\bibitem{GMCASE}
\BIBentryALTinterwordspacing
G.~Motors, ``Path to autonomous,'' Accessed: 2023-4-26. [Online]. Available:
  \url{https://www.gm.com/commitments/path-to-autonomous}
\BIBentrySTDinterwordspacing

\bibitem{9724183}
Z.~Wang, R.~Gupta, K.~Han, H.~Wang, A.~Ganlath, N.~Ammar, and P.~Tiwari,
  ``Mobility digital twin: Concept, architecture, case study, and future
  challenges,'' \emph{IEEE Internet of Things Journal}, vol.~9, no.~18, pp.
  17\,452--17\,467, 2022.

\bibitem{wang2023metamobility}
H.~Wang, Z.~Wang, D.~Chen, Q.~Liu, H.~Ke, and K.~Han, ``Metamobility:
  Connecting future mobility with the metaverse,'' \emph{IEEE Vehicular
  Technology Magazine}, pp. 2--12, 2023.

\bibitem{chen2018digital}
X.~Chen, E.~Kang, S.~Shiraishi, V.~M. Preciado, and Z.~Jiang, ``Digital
  behavioral twins for safe connected cars,'' in \emph{Proc. the 21th ACM/IEEE
  International Conference on Model Driven Engineering Languages and Systems},
  2018, pp. 144--153.

\bibitem{wang2021digital}
Z.~Wang, K.~Han, and P.~Tiwari, ``Digital twin simulation of connected and
  automated vehicles with the unity game engine,'' in \emph{Proc. IEEE 1st
  International Conference on Digital Twins and Parallel Intelligence (DTPI)},
  2021, pp. 1--4.

\bibitem{Nvidia-sim}
\BIBentryALTinterwordspacing
``Nvidia drive sim - powered by omniverse.'' [Online]. Available:
  \url{https://developer.nvidia.com/drive/simulation}
\BIBentrySTDinterwordspacing

\bibitem{wang2022cadsim}
J.~Wang, S.~Manivasagam, Y.~Chen, Z.~Yang, I.~A. B{\^a}rsan, A.~J. Yang, W.-C.
  Ma, and R.~Urtasun, ``{CADS}im: Robust and scalable in-the-wild 3d
  reconstruction for controllable sensor simulation,'' in \emph{6th Annual
  Conference on Robot Learning (CoRL)}, 2022.

\bibitem{mildenhall2021nerf}
B.~Mildenhall, P.~P. Srinivasan, M.~Tancik, J.~T. Barron, R.~Ramamoorthi, and
  R.~Ng, ``Ne{RF}: {R}epresenting scenes as neural radiance fields for view
  synthesis,'' \emph{Communications of the ACM}, vol.~65, no.~1, pp. 99--106,
  2021.

\bibitem{song2021vis2mesh}
S.~Song, Z.~Cui, and R.~Qin, ``Vis2mesh: {E}fficient mesh reconstruction from
  unstructured point clouds of large scenes with learned virtual view
  visibility,'' in \emph{Proc. IEEE/CVF International Conference on Computer
  Vision (CVPR)}, 2021, pp. 6514--6524.

\bibitem{guanetti2018control}
J.~Guanetti, Y.~Kim, and F.~Borrelli, ``Control of connected and automated
  vehicles: {S}tate of the art and future challenges,'' \emph{Annual Reviews in
  Control}, vol.~45, pp. 18--40, 2018.

\bibitem{ye2019evaluating}
L.~Ye and T.~Yamamoto, ``Evaluating the impact of connected and autonomous
  vehicles on traffic safety,'' \emph{Physica A: Statistical Mechanics and its
  Applications}, vol. 526, p. 121009, 2019.

\bibitem{wang2020architectural}
H.~Wang, T.~Liu, B.~Kim, C.-W. Lin, S.~Shiraishi, J.~Xie, and Z.~Han,
  ``Architectural design alternatives based on cloud/edge/fog computing for
  connected vehicles,'' \emph{IEEE Communications Surveys \& Tutorials},
  vol.~22, no.~4, pp. 2349--2377, 2020.

\bibitem{glaessgen2012digital}
E.~Glaessgen and D.~Stargel, ``The digital twin paradigm for future {NASA} and
  {US} {A}ir {F}orce vehicles,'' in \emph{53rd AIAA/ASME/ASCE/AHS/ASC
  Structures, Structural Dynamics and Materials Conference 20th AIAA/ASME/AHS
  Adaptive Structures Conference 14th AIAA}, 2012, p. 1818.

\bibitem{jones2020characterising}
D.~Jones, C.~Snider, A.~Nassehi, J.~Yon, and B.~Hicks, ``Characterising the
  digital twin: A systematic literature review,'' \emph{CIRP Journal of
  Manufacturing Science and Technology}, vol.~29, pp. 36--52, 2020.

\bibitem{liu2022poster}
Y.~Liu, H.~Wang, Z.~Cai, D.~Chen, and K.~Han, ``Poster: Enabling high-fidelity
  and real-time mobility digital twin with edge computing,'' in \emph{Proc.
  IEEE/ACM 7th Symposium on Edge Computing (SEC)}, 2022, pp. 281--283.

\bibitem{kuvsic2023digital}
K.~Ku{\v{s}}i{\'c}, R.~Schumann, and E.~Ivanjko, ``A digital twin in
  transportation: Real-time synergy of traffic data streams and simulation for
  virtualizing motorway dynamics,'' \emph{Advanced Engineering Informatics},
  vol.~55, p. 101858, 2023.

\bibitem{fan2021digital}
B.~Fan, Y.~Wu, Z.~He, Y.~Chen, T.~Q. Quek, and C.-Z. Xu, ``Digital twin
  empowered mobile edge computing for intelligent vehicular lane-changing,''
  \emph{IEEE Network}, vol.~35, no.~6, pp. 194--201, 2021.

\bibitem{deng2022depth}
K.~Deng, A.~Liu, J.-Y. Zhu, and D.~Ramanan, ``Depth-supervised {N}e{RF}: Fewer
  views and faster training for free,'' in \emph{Proc. the IEEE/CVF Conference
  on Computer Vision and Pattern Recognition (CVPR)}, 2022, pp.
  12\,882--12\,891.

\bibitem{mueller2022instant}
T.~M\"uller, A.~Evans, C.~Schied, and A.~Keller, ``Instant neural graphics
  primitives with a multiresolution hash encoding,'' \emph{ACM Trans. Graph.},
  vol.~41, no.~4, pp. 102:1--102:15, Jul. 2022.

\bibitem{li2022rt}
C.~Li, S.~Li, Y.~Zhao, W.~Zhu, and Y.~Lin, ``{RT-NeRF}: Real-time on-device
  neural radiance fields towards immersive {AR/VR} rendering,'' in \emph{Proc.
  the 41st IEEE/ACM International Conference on Computer-Aided Design}, 2022,
  pp. 1--9.

\bibitem{zhang2021ners}
J.~Zhang, G.~Yang, S.~Tulsiani, and D.~Ramanan, ``Ne{RS}: neural reflectance
  surfaces for sparse-view 3{D} reconstruction in the wild,'' \emph{Advances in
  Neural Information Processing Systems}, vol.~34, pp. 29\,835--29\,847, 2021.

\bibitem{wang2021nerf}
Z.~Wang, S.~Wu, W.~Xie, M.~Chen, and V.~A. Prisacariu, ``Ne{RF}--: {N}eural
  radiance fields without known camera parameters,'' \emph{arXiv preprint
  arXiv:2102.07064}, 2021.

\bibitem{xie2023snerf}
Z.~Xie, J.~Zhang, W.~Li, F.~Zhang, and L.~Zhang, ``S-{N}e{RF}: {N}eural
  radiance fields for street views,'' 2023.

\bibitem{guo2020deep}
Y.~Guo, H.~Wang, Q.~Hu, H.~Liu, L.~Liu, and M.~Bennamoun, ``Deep learning for
  3{D} point clouds: A survey,'' \emph{IEEE transactions on pattern analysis
  and machine intelligence}, vol.~43, no.~12, pp. 4338--4364, 2020.

\bibitem{bisheng2017progress}
Y.~Bisheng, L.~Fuxun, and H.~Ronggang, ``Progress, challenges and perspectives
  of 3{D} {LiDAR} point cloud processing,'' \emph{Acta Geodaetica et
  Cartographica Sinica}, vol.~46, no.~10, p. 1509, 2017.

\bibitem{martinbrualla2021nerf}
R.~Martin-Brualla, N.~Radwan, M.~S. Sajjadi, J.~T. Barron, A.~Dosovitskiy, and
  D.~Duckworth, ``Ne{RF} in the wild: Neural radiance fields for unconstrained
  photo collections,'' in \emph{Proc. the IEEE/CVF Conference on Computer
  Vision and Pattern Recognition (CVPR)}, 2021, pp. 7210--7219.

\bibitem{oughton2019open}
E.~J. Oughton, K.~Katsaros, F.~Entezami, D.~Kaleshi, and J.~Crowcroft, ``An
  open-source techno-economic assessment framework for 5{G} deployment,''
  \emph{IEEE Access}, vol.~7, pp. 155\,930--155\,940, 2019.

\bibitem{xiao2021pandaset}
P.~Xiao, Z.~Shao, S.~Hao, Z.~Zhang, X.~Chai, J.~Jiao, Z.~Li, J.~Wu, K.~Sun,
  K.~Jiang \emph{et~al.}, ``Pandaset: Advanced sensor suite dataset for
  autonomous driving,'' in \emph{Proc. IEEE International Intelligent
  Transportation Systems Conference (ITSC)}, 2021, pp. 3095--3101.

\end{thebibliography}

\end{document}